\documentclass[twocolumn,aps]{revtex4}
\usepackage{amsmath}
\usepackage{graphicx}
\usepackage{amssymb}

\makeatletter

\usepackage{graphicx}

\makeatother
\begin{document}

\title{Spin-charge separation in ultra-cold quantum gases }

\author{A. Recati,$^{1)}$ P.O. Fedichev,$^{1)}$ W. Zwerger$^{2)}$, P.
Zoller$^{1)}$}

\affiliation{$^{1}$ Institute for Theoretical Physics, University of Innsbruck,
A--6020 Innsbruck, Austria.}

\affiliation{$^{2}$ Sektion Physik, Universit\"at M\"uunchen, Theresienstr. 37/III, D-80333 M\"unchen,
Germany.}

\begin{abstract}
We investigate the physical properties of quasi-1D quantum gases of
fermion atoms confined in harmonic traps. Using the fact that for
a homogeneous gas, the low energy properties are exactly described
by a Luttinger model, we analyze the nature and manifestations of
the spin-charge separation. Finally we discuss the necessary physical
conditions and experimental limitations confronting possible experimental
implementations.
\end{abstract}
\maketitle

One dimensional (1D) quantum liquids are very rich and interesting
systems. In spite of their apparent conceptual simplicity, both the
ground state and the excitations exhibit strong correlation effects
and posses a number of exotic properties, ranging from spin-charge
separation to fractional statistics (see \cite{luttinger:haldane,schulz:review,lutt:fraqreview}
and ref. therein). Progress in creating, manipulating and studying
ultra-cold quantum gases with controlled and adjustable interactions
\cite{BEC,Pitaevskii:review}, and in particular the recent
development of 1D magnetic and optical wave guides opens the door
for a new and clean physical realization of such 1D systems with the
tools of atomic physics and quantum optics. While most of the recent
theoretical and experimental work has focused on 1D Bose gases (as a Tonks gas or a
quasicondensate) \cite{salamon2002,petrov:1Dinteraction} progress in cooling Fermi gases into the quantum
degenerate regime \cite{expfermi} point to the possibility of realizing a Luttinger
Liquid (LL) \cite{schulz:review} with cold fermionic atoms. One of the key predictions
of the Tomonaga-Luttinger model (LM) for interacting fermions is spin-charge
separation \cite{schulz:review}. It is a feature of interacting spin-$1/2$
particles and manifests itself in complete separation in the dynamics
of spin and density waves. Both branches of the excitations are sound-like
and characterized by different propagation velocities. This phenomena
is one of hallmarks of a Luttinger liquid, however it has never been
been demonstrated in a clean way in an actual condensed matter system
(see e.g. \cite{lutt:spinchargeobserve1,lutt:spinchargeobserve2}).
It is the purpose of this Letter to analyze in detail the conditions
of realizing an (inhomogeneous) LL with a gas of cold fermionic atoms
in 1D harmonic trap geometries, and in particular to study the possibilities
of seeing spin-charge separation in the spectroscopy and wave packet
dynamics of laser excited 1D Fermi gases.

\begin{figure}
\includegraphics[  width=0.90\columnwidth,
  keepaspectratio]{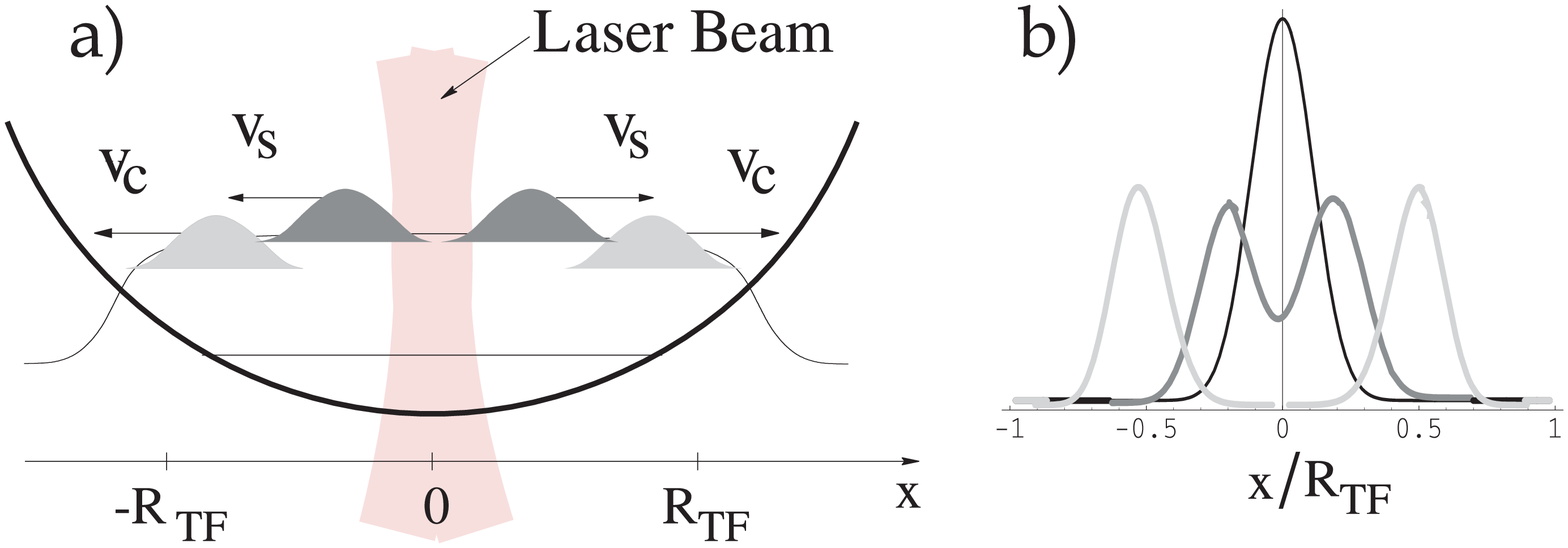} Fig.~1. (a) Schematic setup: a two component fermi gas is trapped
in a harmomic potential in a 1D configuration. At time $t=0$ a
short laser pulse focused near the center of the trap excites (i)
a density (charge) or (ii) a spin-wavepacket. (b) Wave packet
dynamics for different times as function of position (in units of
the Thomas Fermi radius $R_{TF}$): spin-charge separation
manifests itself in a spatial separation of the spin (solid line)
and density (dashed line) wave packets (shown at half a trap
oscillation period $\omega t=0.5$), which can be probed by a
second short laser pulse at a later time. The parameters
correspond to $N=10^3$ $^6$Li atoms in a trap with $\omega=1$Hz
(with coupling parameter $\xi=1$, see text).
\end{figure}

The simplest example of a Luttinger liquid made of a gas of cold
atoms consists of fermionic atoms with two ground states
representing a spin-$1/2$ under quasi-1D trapping conditions. We
assume the atoms to be cooled below the Fermi-degeneracy
temperature $k_{B}T_{F}\sim N\hbar \omega $, where $N$ is the
number of particles and $\omega $ is the frequency of the
longitudinal confinement. The condition for a quasi-1D system is
tight transverse trapping in an external potential with the
frequency $\omega _{\perp }$ exceeding the characteristic energy
scale of the longitudinal motion. Due to the quantum degeneracy
the longitudinal motion has all the energy levels up to the
Fermi-energy $\epsilon _{F}\sim k_{B}T_{F}$ filled. Thus we
require the total number of particles to be restricted by $N\ll
\omega _{\perp }/\omega $, which for realistic traps is of the
order of a few hundred or thousand. Let us now turn to an estimate
of the conditions to reach the strongly interacting limit. Since
the interaction between the atoms has a range much smaller than
the interparticle spacing, at low temperatures only collisions
between particles with different spins are allowed by the
exclusion principle. Therefore, all the relevant interactions are
characterized by a single parameter, the scattering length $a$
corresponding to inter-component interaction. The effective 1D
interaction can thus be represented as a zero-range potential of
the strength $g=2\pi \hbar ^{2}a/ml_{\perp }^{2}$, where $l_{\perp
}$ is the width of the ground state in the transverse direction
($a\ll l_{\perp }$)\cite{petrov:1Dinteraction} and $m$ is the mass
of the gas particle. The interaction strength in a Luttinger
liquid is then characterized by the dimensionless parameter $\xi
=g/\pi \hbar v_{F}$, where $mv_{F}^{2}/2=k_{B}T_{F}$ is the Fermi
velocity (at trap center). Remarkably, in a trapped gas $\xi \sim
(a/l_{\perp })(\omega _{\perp }/\omega N )^{1/2}$ and thus can be
tuned externally either by changing the transverse confinement, or
by changing the scattering length by magnetic field. The ratio of
the transverse and the longitudinal frequencies is quite large, we
can easily reach the strong coupling limit $\xi\sim 1$, even in a
dilute gas.

Below we will study spin-charge separation according to the
schematic setup outlined in Fig.~1. We assume that a short far
off-resonant laser pulse is focused at the center of the harmonic
trap with a two-component atomic Luttinger liquid, where depending
on the laser parameters (e.g. light polarization) density and spin
wave packets can be excited (Fig.~1a). Spin-charge separation
manifests itself in different propagation speed of the spin and
density wave packets (see Fig.~1b). This can be probed at a later
time with a second short laser pulse. Spin dependent optical
potentials can be generated by a laser tuned e.g. between fine
structure levels of excited Alkali states acts on the ground state
``spin'' in a way equivalent to the external magnetic field
interacting with the spin density and thus introduce a spin
density perturbation. The goal of the following derivations is
thus to (i) derive the frequencies of the spin and charge modes of
atomic Luttinger liquid confined in a harmonic trap, and (ii) to
discuss the wavepacket dynamics as superposition of these modes.

We analyze the properties of a trapped Luttinger liquid by
combining Haldane's low energy hydrodynamic description
\cite{luttinger:haldane} with a local density approximation. As
was shown by Haldane, any homogeneous interacting quantum liquid
in one dimension can be described by an affective hydrodynamic
Hamiltonian, which completely describes the behavior at
wavelengths much larger than the interparticle spacing. In the
case of spin-$1/2$ fermions, the effective Hamiltonian reads
\begin{equation} H=\sum _{\nu =\rho ,s}\int dx\frac{\hbar v_{\nu
}}{2}\left[K_{\nu }\Pi _{\nu }^{2}+\frac{1}{K_{\nu }}(\partial
_{x}\phi _{\nu })^{2}\right],\label{spinchargeHam}\end{equation}
 Here the index $\nu $ counts the spin ($s$) and the density ($\rho $)
excitations. The phenomenological parameters $K_{\nu }$ and $v_{\mu }$
completely characterize the low energy physics. For exactly solvable
models like the 1D lattice model, can be directly expressed in terms
of the microscopic parameters of the theory \cite{schulz:hubbard}.
The gradients of the phases $\partial \phi _{\nu }$ are the density
and the spin density fluctuations respectively. The canonical momenta
$\Pi _{\nu }$ conjugated to the phases $\phi _{\nu }$ are related
to the spin and the density currents $j_{\nu }=v_{\nu }K_{\nu }\Pi _{\nu }$.
In a rotationally invariant Fermi-gas the quantity $K_{s}=1$, so
that only independent parameters are $K_{\rho }$, and $v_{s,\rho }$\cite{luttinger:haldane,schulz:review}.

As apparent from Eq.~(\ref{spinchargeHam}) a distinctive feature
of the LM Hamiltonian is the complete separation of the spin and the
{}``charge'' degrees of freedom. In a spatially homogeneous gas
the spin and the charge waves propagate at the velocities $v_{c}$
and $v_{s}$ respectively. This is the essence of the spin-charge
separation phenomena. In our cold atoms system the charge and the
spin waves are modeled by excitations of the total and the relative
densities of the components.

The local density approach to model a trapped gas assumes that the
size of the atom cloud $R\gg k_{F}^{-1}$, i.e. the size of the gas
sample is much larger than the interparticle separation, consistent
with $N\gg 1$. The variation of $K_{\mu }$ and $v_{\mu }$ is assumed
to originate only from the spatial dependence of the gas density,
$K_{\mu }[n]\rightarrow K_{\mu }[n(x)]$ and $v_{\mu }[n]\rightarrow v_{\mu }[n(x)]$.
We will also assume that the numbers of particles of the both ``spins''
are the same (i.e. the total spin of the system is zero). Our goal
is now to derive the excitation spectrum in the charge and the spin
sector analytically in the weak and strong coupling limits, and study
the intermediate regime using numerical techniques. As a first step,
this requires the calculation of the ground state density distribution.

Within the local density approximation, the ground state of the system
can be characterized using the Thomas-Fermi equilibrium condition:
\begin{equation}
\frac{dE(n)}{dn}=\mu -V(x),\label{TF}\end{equation}
 Here $E(n)$ is the internal energy of the gas per unit length of
the gas as a function of its (total) density, $\mu $ is the chemical
potential, $V(x)=m\omega ^{2}x^{2}/2$ is the longitudinal external
potential, and $\omega $ is the frequency of the longitudinal confinement.
This equation is just the expression of the fact that the energy cost
of adding a particle to the system equals to the chemical potential
corrected by the local value of the external potential. Generally
speaking, the external potentials acting on the two different \char`\"{}spin\char`\"{}
components can be different. This feature can be used to generate
offsets in the densities of the components and hence produce the spin
and the density excitations with laser light (see below).

In preparation for the interacting case we consider first a free gas
ground state ($g=0$). The density of the gas at a given position
$x$ is related to the local value of Fermi momentum, $k_{F}(x)=\pi n(x)/2$.
The internal energy of the gas is just the density of the kinetic
energy (the so called quantum pressure) $E(n)=\hbar ^{2}\pi ^{2}n^{3}(x)/24m$.
Substituting these expressions into Eq.(\ref{TF}) we find \begin{equation}
n_{TF}(x)=n_{0}\sqrt{1-\frac{x^{2}}{R_{TF}^{2}}},\label{nTF}\end{equation}
 for $|x|<R_{TF}$, and $0$ otherwise. Here $n_{0}\equiv n(x=0)=(8\mu m/\hbar ^{2}\pi ^{2})^{1/2}$
is the density in the center and $R_{TF}=(2\mu /m\omega ^{2})^{1/2}$
is the Thomas-Fermi size of the cloud \cite{fermi:idealhydro}. From the requirement that the integrated density equals the particle
number we have the condition $\mu =\hbar \omega N/2$.

The excitations of the gas can be found from the equations
of motion following from the Hamiltonian (\ref{spinchargeHam}): \begin{equation}
\dot{\phi }_{\nu }=K_{\nu }(x)v_{\nu }(x)\Pi _{\nu },\, \dot{\Pi }_{\nu }=\frac{\partial }{\partial x}\frac{v_{\nu }(x)}{K_{\nu }(x)}\frac{\partial }{\partial x}\phi _{\nu }.\label{bogdegen}\end{equation}
 In particular, for an ideal gas we can use the density profile (\ref{nTF})
and find the equations for the mode functions: \[
-\epsilon ^{2}\phi _{\nu }=\hbar ^{2}\omega ^{2}(1-\tilde{x^{2}})^{1/2}\frac{\partial }{\partial \tilde{x}}(1-\tilde{x^{2}})^{1/2}\frac{\partial }{\partial \tilde{x}}\phi _{\nu },\]
 where $\tilde{x}=x/R_{TF}$. The solution is given by $\phi _{\nu n}=A_{\nu }\sin (\epsilon _{n}\arccos \tilde{x})+B_{\nu }\cos (\epsilon _{n}\arccos \tilde{x})$.
The discrete spectrum of eigenfrequencies is found by analyzing the
boundary conditions: $\epsilon _{\nu }(n)=\hbar \omega (n+1)$ \cite{fermi:idealhydro}
both for the spin and the density modes. The first modes ($n=0$)
and the wavefunctions $\phi _{\nu }\sim x/\sqrt{1-\tilde{x}^{2}}$
correspond to harmonic oscillations of the center of mass of the total
density and the total spin (dipole modes).

Before starting with the perturbation theory in small interaction
parameter $\xi \ll 1$, we note that in a finite system there is additional
energy scale, which is the level spacing ($\hbar \omega $). In order
for the interaction effects manifest themselves in a way similar to
a bulk system, we need the interaction to be stronger than the level
spacing, $n(x)g\gg \hbar \omega $, in contrast to the case of very weak interactions studied in \cite{Wonneberger}.

In a homogeneous gas the Luttinger parameters in the Hamiltonian (\ref{spinchargeHam})
to the lowest order in $\xi =g/\pi \hbar v_{F}\ll 1$ can be found
using perturbation theory: $K_{s}=1$, $v_{s}=v_{F}(1-\xi /2)$, $K_{\rho }=1-\xi /2$
and $v_{\rho }=v_{F}(1+\xi /2)$ \cite{lutt:review}. The energy of
the ground state can be obtained by averaging the interparticle interaction
over the ground state, $E_{0}=\hbar ^{2}\pi ^{2}n^{3}/24m+gn^{2}/4$.
In the spirit of the local density approximation we substitute a spatially
dependent density $n(x)$ in the expressions for homogeneous gas.
Then, using Eq.(\ref{TF}) we find that in the first order in $\xi =g/\pi v_{F}(0)$,
the density of the gas uniformly decreases by $\delta n(x)=-2gm/\hbar ^{2}\pi ^{2}$,
i.e. the interaction reduces the density, as expected. This simple
conclusion holds everywhere as long as $g\ll \hbar ^{2}n(x)/m$, i.e.
\begin{equation}
(R_{TF}-x)/R_{TF}\alt (gm/\hbar ^{2}n_{0})^{2}\sim O(\xi ^{2})\label{border}\end{equation}
 The velocities of the spin and the density waves are \begin{equation}
v_{s,\rho }(x)=\frac{\pi \hbar n_{TF}(x)}{2m}(1-\frac{A_{s,\rho }gm}{\pi ^{2}\hbar ^{2}
n_{TF}(x)}),\label{eq:velocities}\end{equation}
 where $A_{s}=3$ and $A_{\rho }=1$. Using the expansion in powers
of $\xi $ of the Luttinger parameters and the density profile (\ref{TF})
we find, that the frequency of the density dipole mode does not depend
on the interaction (as it should be), while the the spin dipole mode
shift is given by an integral logarithmically diverging at the border
of the gas cloud. The divergence occurs due to localization of the
excitations of a free gas close to the gas cloud border and arises
in any potential, which is a power law in $x$. Using the condition
(\ref{border}) to cut off the divergence, we find \[
\delta \omega _{s1}=-\omega \frac{3gm}{\pi ^{2}\hbar ^{2}n_{TF}(0)}\log \frac{\pi \hbar ^{2}n_{TF}(0)}{2gm}.\]
 This shift is negative and can be observed by comparing the spin
and the density oscillations of the gas cloud. Note that in a harmonic
trap the perturbation theory requirement is stronger than in a homogeneous
Luttinger liquid: we have to require $\xi\log (1/\xi)\ll 1$
instead of simply $\xi\ll 1$.

For higher modes the application of the perturbation theory in Eqs.(\ref{bogdegen})
turns inconvenient and the frequencies of the excitations can be analyzed
within the WKB approximation. The accuracy of the WKB spectrum estimation
is $\sim 1/\pi ^{2}n^{2}$ \cite{migdal:quant}, whereas the expected
corrections are of order $g/v_{F}.$ Therefore for sufficiently high
$n$ the eigenfrequencies can be reliably obtained from the WKB quantization
condition\cite{LL:volIII}, \begin{equation}
\int _{-x_{0}}^{x_{0}}p(x)dx=\hbar \pi (n+\alpha ),\label{bornsommer}\end{equation}
 where $p(x)$ is the WKB momentum corresponding to a given energy,
$n$ is the (integer) quantum number, $x_{0}$ is the classical turning
point and the constant $\alpha =1$ is fixed by comparing the WKB results and the
exact solutions of Eqs.(\ref{bogdegen}) for a weakly interacting
gas. Substituting the dispersion relation $\epsilon =v_{\rho ,s}(x)p(x)$
with the velocities (\ref{eq:velocities}) into Eq.(\ref{bornsommer})
we obtain the the same sort of logarithmically diverging integrals
as those for in the perturbation theory above. By regularizing them
using the condition (\ref{border}) we find, that \begin{equation}
\epsilon _{\rho ,s}=\hbar \omega (n+1)(1-\frac{2gmA_{\rho ,s}}{\pi ^{2}\hbar ^{2}n_{TF}(0)}\log (\frac{\hbar ^{2}n_{TF}(0)\pi }{mg})).\label{eq:spectrum}\end{equation}
 This simple WKB calculation confirms the interaction dependent split
of the spin and the density oscillation frequencies. %
\begin{figure}
\includegraphics[  width=0.90\columnwidth,
  keepaspectratio]{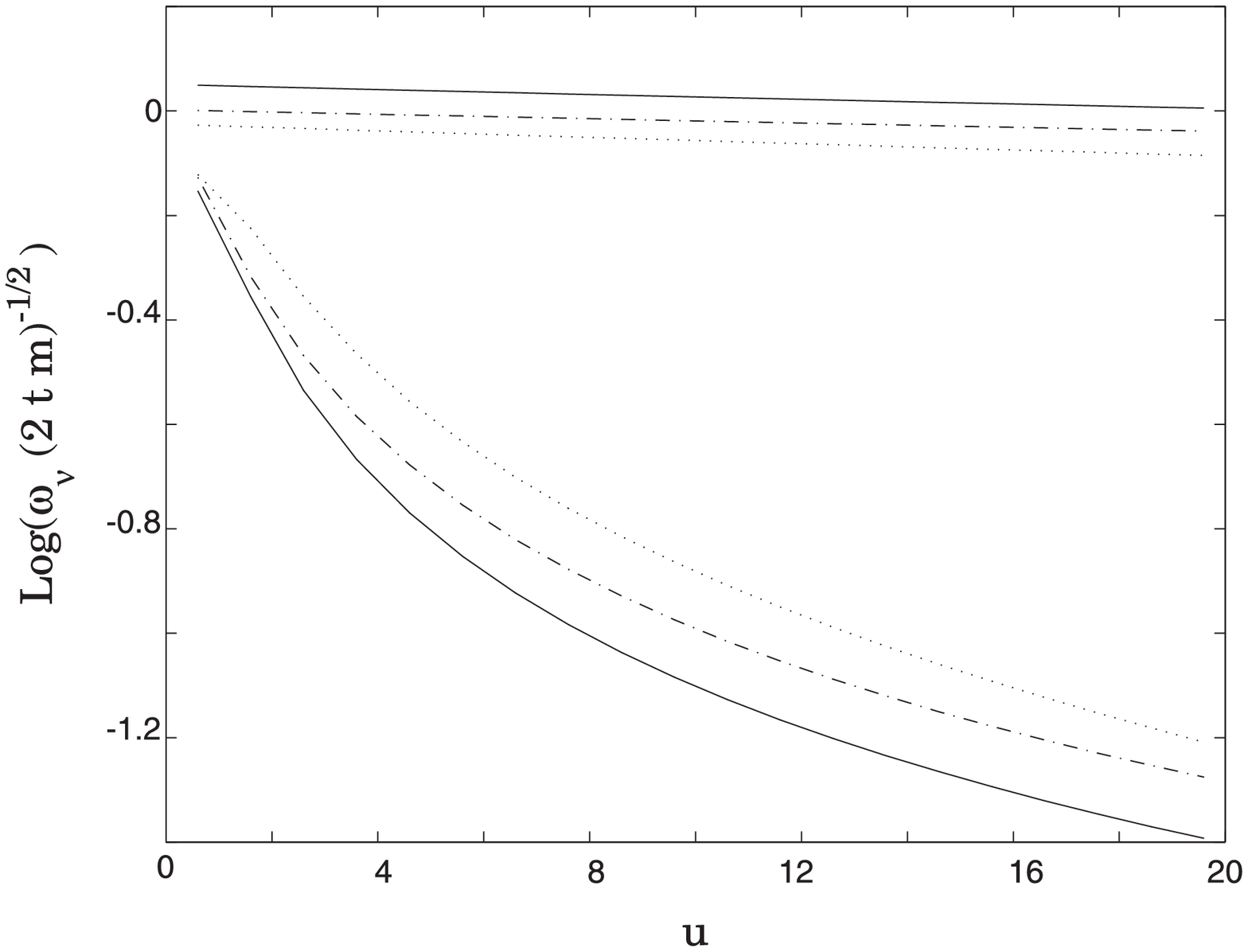}
Fig.~3 The level spacing (in units of $\omega$) between the spin
(solid line) and the ``charge'' (dashed line) modes vs. $\log\xi$
for different central densities: $n(0)=0.25$ (solid), $n(0)=0.42$
(dot-dashed), $n(0)=0.58$ (dotted). \vspace{-0.5cm}
\end{figure}

In the limit of very large interaction strength ($g\gg \pi \hbar v_{F}$)
the repulsion between the atoms of the two different species is very
strong. Hence, the properties of the gas are similar to those of an
ideal single component gas of indistinguishable particles. The density
profile is still given by Eq.(\ref{nTF}), but now with $n_{0}=n_{\infty }=(2\mu m/\hbar ^{2}\pi ^{2})^{1/2}$,
$\mu =\hbar N\omega $, and $R_{TF}=R_{\infty }=(2\mu /m\omega ^{2})^{1/2}$.
This distribution is less dense and thus broader than that for a weakly
interacting gas. The density wave speed is equal to the Fermi velocity
$v_{F}=\hbar \pi n_{\infty }/m$ and, after integration in Eq.(\ref{bornsommer}),
we find that spectrum of the density waves is the same as in the non-interacting
case above. In turn, the relation between the energy and the WKB momentum
for the spin wave is given by \[
\epsilon _{s}(p)=\frac{B\hbar n^{2}(x)p(x)}{m^{2}g},\]
 where the coefficient $B\sim 13$ $(\pm 2)$ was found numerically. Once again, using the quantization condition
(\ref{bornsommer}) and cutting off the logarithmically divergent
integral at the point $n\ll g$, we find, that \[
\epsilon _{ns}=\hbar \omega (n+\alpha )\frac{\hbar ^{2}Bn_{\infty }}{gm\log (gm/\hbar ^{2}n_{\infty })},\]
 where $\alpha \sim 1$. As it is clear from the latter expression,
the interaction profoundly changes the properties of the spin mode.
In the limit of the strong interaction the level spacing decreases
and is much smaller than that between the density waves ($\omega $).

In order to confirm our analytical results, we performed a
numerical calculation valid for arbitrary interaction strength
based on a lattice model. Using the exact solution
\cite{lieb:fermions,lutexcit:coll} for calculation of the
Luttinger constants and the Thomas-Fermi approximation (\ref{TF}),
we determined the WKB level spacings for the spin and the charge
modes. The results are presented in Fig.~2 as a plot of the
excitations level spacing vs. the dimensionless interaction
strength $g/\pi v_{F}$ calculated at the center of the trap.

As outlined in Fig.~1, wavepackets of the spin and density
excitations can be generated by short off-resonant and state
selective laser pulses focused to a spot size $\ell $ with $R\gg
\ell \gg k_{F}^{-1}$, where $R_{TF}$ is the size of the atom cloud
and $k_{F}^{-1}$ the interparticle distance. This procedure is
analogous to the MIT setup originally used to study propagation of
sound waves in elongated condensates \cite{MIT:soundprop}. Fig.~1b
shows as an example the wave packet dynamics for the states
$|F=1/2, M_F=\pm 1/2\rangle$ of $^{}6$Li with interaction
parameter $\xi=1$, corresponding to $N=500$ particles at trap
frequency of $\omega=1$Hz, $\omega_\perp=250$kHz and scattering
length $a_s= 23 \AA$. Tuning near the Feshbach resonance (at
$B=800$G ) allows an increase of the scattering length by one
order magnitude, allowing for $N=1000$ atoms at a trap frequency
of $\omega_\perp=100$kHz.

Let us finally now turn to a discussion of the life time of the
excitations and temperature requirements. The Hamiltonian
(\ref{spinchargeHam}) represents only the first term in
hydrodynamic expansion $q/k_{F}\ll 1$. The higher order terms
originate from, for example, non-linearity of the fermionic
spectrum and mix the excitations with each other. The first
corrections are of the third power in $\Pi $ and $\partial \phi $,
and hence lead to scattering of the excitations. To study the
relaxation phenomena we switch for simplicity to a case of a
single-component LM. According to Haldane, the generic term is
given by $V_{int}=\gamma \hbar ^{2}/m (\partial \phi )^{3}$, where
$\gamma \sim 1$ \cite{luttinger:haldane}. The damping of the
oscillations can be found from Fermi's Golden rule applied to the
interaction Hamiltonian $V_{\textrm{int}}$ and using the second
quantized representation for the phonon field.A straightforward
calculation gives $\Gamma _{T=0}\sim \hbar ^{2}q^{4}L/m^{2}u$. for
$k_{B}T\ll \hbar \omega _{q}$, which is the decay rate in a
process where a particle with the energy $\hbar \omega _{q}$
decays into a pair of particles with $\omega _{1},\omega
_{2}<\omega _{q}$). Note that $\omega _{q}\gg \omega $, otherwise
there are no final states for such decay instability (i.e. the
lowest excitations are very stable at very low temperatures). In
the case of $k_{B}T\gg \hbar \omega _{q}$(but still $T\ll T_{F}$)
we find $\Gamma _{T}\sim q^{2}Lk_{B}^{2}T^{2}/m^{2}u^{3}$ which
corresponds to Landau damping, i.e. the contribution of a process
in which the damping occurs by scattering a high-frequency
excitation with $\hbar \omega _{q}^{\prime }\sim k_{B}T$. Both
results contain the size of the sample $L$ and are only valid in
the collisionless regime $\Gamma /\omega _{q}\ll 1$. This is the
case for sufficiently small temperatures (or high number of
particles). Indeed, for the lowest excitations $Lmu/\hbar \sim N$,
$\omega _{q}\sim \omega \sim \epsilon _{F}/\hbar N$, so that
$\Gamma /\omega _{q}\sim (k_{B}T/\epsilon _{F})^{2}\ll 1$,
i.e.~the excitations are only weakly damped.

In conclusion we performed the analysis of a double component Fermi
gas confined in a harmonic trap. Based on the LM we have investigated
the nature of the excitations and analyzed an experiment where spin-charge
separation can be observed {}``directly'' in experiments addressing
the spectral properties of the lowest excitations with laser light.

Discussions with J.I. Cirac, J. von Delft, D. Jaksch and
U. Schollw\"ock are gratefully acknowledged. Work supported in part by the
Austrian Science Foundation and EU Networks, and the Institute
for Quantum Information.

\end{document}